\documentclass[12pt]{article}
\usepackage{latexsym,amsmath,amsfonts,amssymb}
\usepackage[latin1]{inputenc}
\usepackage[pdftex]{graphicx}

\usepackage{euscript,mathrsfs}
\usepackage{fullpage}
\usepackage{color}
\usepackage{bbm}
\usepackage[american]{babel}

\makeatletter \@addtoreset{equation}{section}

\makeatletter\renewcommand\section{\@startsection {section}{1}{\z@}%
                                   {-3.5ex \@plus -1ex \@minus -.2ex}%nn
                                   {2.3ex \@plus.2ex}%
                                   {\normalfont\large\bfseries}}
\renewcommand\subsection{\@startsection{subsection}{2}{\z@}%
                                     {-3.25ex\@plus -1ex \@minus -.2ex}%
                                     {1.5ex \@plus .2ex}%
                                     {\normalfont\bfseries}}

\renewcommand{\baselinestretch}{1.2}
\newcommand{\be}{\begin{equation}}
\newcommand{\ee}{\end{equation}}
\newcommand{\bea}{\begin{eqnarray}} 
\newcommand{\eea}{\end{eqnarray}}

\newcommand{\comment}[1]{}

\newcommand{\nn}{\nonumber}

\newcommand{\p}{\partial}

\newcommand\rref[1]{(\ref{#1})}

\newcommand{\Z}{\mathbb{Z}}

\newcommand{\M}{{\cal M}}

\begin{document}

\begin{titlepage}
\vspace{6cm}
\vfil\

\begin{center}
{\LARGE Quantum Topologically Massive Gravity in de Sitter Space} 
%\end{center}

\vspace{6mm}
%\begin{center}
{Alejandra Castro\footnote{e-mail: {\tt acastro@physics.mcgill.ca}}, %$^{a}$, 
 Nima Lashkari\footnote{e-mail: {\tt lashkari@physics.mcgill.ca}}, %$^{a}$ 
 \&
Alexander Maloney\footnote{e-mail: {\tt maloney@physics.mcgill.ca}}%$^{a}$
}\vspace{6.0mm}\\
\bigskip%\medskip
%\smallskip\centerline{$^a$ \it 
{\it 
McGill Physics Department, 3600 rue University, Montreal, QC H3A 2T8, Canada}
%\medskip
\vfil

\end{center}
\setcounter{footnote}{0}

\begin{abstract}
\noindent
We consider three dimensional gravity with a positive cosmological constant and non-zero gravitational Chern-Simons term.  This theory has inflating de Sitter solutions and local metric degrees of freedom.  The Euclidean signature partition function of the theory is evaluated including both perturbative and non-perturbative corrections.  The perturbative one-loop correction is computed using heat kernel techniques.  The non-perturbative corrections come from gravitational instantons with non-trivial topology which can be enumerated explicitly.  We compute the sum over an infinite class of geometries and show that, unlike the case of pure Einstein gravity, the partition function is finite.  This demonstrates that the inclusion of non-trivial local degrees of freedom can render the sum over geometries convergent.
\end{abstract}
%%%%%%%%%%%%%%%%%%%%%%%%%%%%%%%%%%%%%%%%%%%%%%%%%%%%%%%%%%%%%%%%%%%%%%%%%%%%%%%%%%%%%%%%%
\vspace{0.5in}

\end{titlepage}
\newpage

\renewcommand{\baselinestretch}{1.1}  %looks better
\renewcommand{\arraystretch}{1.5}

\tableofcontents

\section{Introduction and Discussion}

Quantum gravity in de Sitter space is a  notoriously difficult subject.  Many basic questions -- such as those involving the nature of observables in de Sitter space and the origin and interpretation of  de Sitter entropy -- remain unresolved.   Many techniques which have proven useful in other circumstances, such as those involving  AdS/CFT and supersymmetry, do not address these issues.
It is natural therefore to focus on  simple theories of de Sitter  gravity where the deep questions of quantum cosmology can be discussed in a quantitative manner.  

Recently, we have focused on the question of three dimensional Einstein gravity with a positive cosmological constant  \cite{Castro:2011xb}.  Although this theory contains inflating de Sitter solutions it has no local degrees of freedom.  This allowed us to perform a series of precise computations which are impossible in more complicated theories of de Sitter gravity.
We were able to compute the exact partition function in Euclidean signature,
which is schematically  written as a path integral of the form
\be\label{aa}
Z= \int {\cal D}g \, e^{-S[g]}~,
\ee
where $S[g]$ is the Einstein-Hilbert action.
The physical interpretation of this path integral is the following.  Euclidean path integrals are used to define states of the Lorentzian theory; the sum over compact geometries with specific boundary data defines the  ``wave function of the universe" in the sense of Hartle and Hawking \cite{Hartle:1983ai}.  The path integral \rref{aa} is an integral over compact manifolds without boundary and it is interpreted as the norm of the Hartle-Hawking wave function.

Of course, functional integrals such as \rref{aa} cannot usually be defined precisely.  However, one can give a precise definition to a path integral if one is able to
\begin{itemize}
\item
Ennumerate all  solutions to the classical equations of motion, which appear as saddle point contributions to \rref{aa}
\item
Compute the infinite series of quantum corrections around each saddle point
\item
Perform the sum over saddle points, including this infinite series of quantum corrections.
\end{itemize}
All three of these computations can be performed explicitly in the case of three dimensional Einstein gravity in de Sitter space.   

In fact, the first two of these tasks are not particularly difficult.  The classification of solutions is related to the classification of spherical three-manifolds and proceeds much like the classification of crystallographic groups in three dimensions.  The quantum corrections can be computed using the relationship between Einstein gravity and Chern-Simons theory.  It is the third task -- the sum over geometries -- which proved most problematic.  This sum is over an infinite number of topologically distinct geometries, and turns out to diverge in a way which cannot  be cured using standard regularization techniques.  Thus the Hartle-Hawking state of Einstein gravity is non-normalizable.

In this paper we consider the path integral \rref{aa} for a slightly different theory of gravity, that of Einstein gravity with a positive cosmological constant and a gravitational Chern-Simons term.  The classical solutions of the theory include the usual inflating de Sitter solutions of Einstein gravity with a positive cosmological consant.  However, the Chern-Simons Lagrangian is third order in derivatives so the theory now possesses a local degree of freedom.  This theory is known as Topologically Massive Gravity (TMG) and was first considered in \cite{Deser:1982sv,Deser:1981wh,Deser:1982vy}.  
Despite the presence of this local degree of freedom, this partition function can be studied in considerable detail.  In this paper we will  focus primarily on those aspects of the computation which differ from the Einstein gravity case considered in \cite{Castro:2011xb}.  Although this paper is self-contained we will quote certain results from \cite{Castro:2011xb}.

The classical sum over geometries in TMG is described in section 2.  The classification of Euclidean geometries  is in fact identical to that of Einstein gravity.  However, the classical action of TMG evaluated on one of these solutions differs from that of Einstein gravity.  The action is complex in Euclidean signature, so that each saddle point geometry is now weighted by a phase.  If the Chern-Simons coupling is appropriately quantized these phases  lead to cancelations in the sum over geometries.  This has the effect of making the sum over geometries more convergent than in  Einstein gravity.

We then turn to the computation of quantum corrections, which are considered in section 3.  This is somewhat more complicated than in the Einstein gravity case, as the theory  possesses a local degree of freedom.   Nevertheless, an explicit computation of the one loop determinant is possible.  This is accomplished by using heat kernel techniques to compute the spectrum of the massive graviton wave operator and zeta function regularization to compute the regulated functional determinant.  It also involves a careful treatment of the gauge fixing terms and Fadeev-Popov ghosts.   
In  appendix A we give a careful derivation of the one-loop determinant of TMG using BRST quantization.  In appendix B we give a detailed analysis, including a numerical study, of the resulting one loop determinant.

Our conclusion is that, unlike the case of Einstein gravity, the sum over geometries converges when the gravitational Chern-Simons coupling takes certain discrete values.  In this computation we include only those physically motivated saddles which have a natural interpretation in Lorentzian signature; this will be discussed in more detail in section 2.  The inclusion of other saddles was considered in the case of Einstein gravity and shown not to qualitatively effect the result.  We expect the same to be true here.
We also include only the one-loop perturbative correction.  Higher loop corrections are certainly present, and indeed quite difficult to compute in TMG.  Our expectation  is that these higher corrections will not effect our conclusions.  This is based on our experience with Einstein gravity, where these higher order corrections were computed explicitly and shown not to alter the divergence structure of the sum over geometries  \cite{Castro:2011xb}.

We emphasize that our conclusion -- that the sum over geometries converges only when a gravitational Chern-Simons term is included -- is very similar to the corresponding result in Anti-de Sitter space.  In that case the partition function of three dimensional Einstein gravity with a negative cosmological constant can be computed exactly, but the result does not have a quantum mechanical interpretation  \cite{Maloney:2007ud}.  Once a gravitational Chern-Simons term is added, however, the sum over geometries has a natural quantum mechanical interpretation as the partition function of a dual conformal field theory \cite{Li:2008dq, Maloney:2009ck}.  We are finding a similar result in the case of a positive cosmological constant.  This may indicate that pure quantum gravity in de Sitter space makes sense only when an appropriate gravitational Chern-Simons term is included.  This is likely to have implications for the conjectured dS/CFT correspondence \cite{Strominger:2001pn}; see e.g. \cite{Anninos:2009jt, Anninos:2011vd} for related considerations. We hope to return to this in the future.

Finally, we  note that other  modifications of  Einstein gravity to include additional degrees of freedom may lead to similar results.  Other straightforward extensions of three dimensional gravity include supersymmetric theories \cite{Achucarro:1987vz,Achucarro:1989gm},  generalized massive gravity \cite{Bergshoeff:2009hq,Bergshoeff:2009aq} and  higher spin theories \cite{Blencowe:1988gj,Bergshoeff:1989ns}.  It would be interesting to compute the partition function in these cases as well.

%%%%%%%%%%%%%%%%%%%%%%%%%%%%%%%%%%%%%%%%%%%%
%%%%%%%%%%%%%%%%%%%%%%%%%%%%%%%%%%%%%%%%%%%%
\section{The Classical Analysis}

In this section we study the partition function at the classical tree level approximation.  In this limit the partition function \eqref{aa} is given by its saddle point approximation
\be
Z=\sum_{g_c}e^{-S^{0}[g_c]+\ldots}~.
\ee
Here the sum is over all classical solutions $g_c$ to the Euclidean equations of motion and $S^{0}[g_c]$ is the classical action.  We will start by identifying the classical solutions $g_c$ and describing their physical interpretation.  We evaluate the tree level action $S^0$ for TMG.  We then explicitly perform the sum over geometries, including an infinite class of saddles with a clear physical interpretation in Lorentzian signature. 

\subsection{Classical solutions}\label{sec:solutions}

Topologically Massive Gravity (TMG) is three dimensional general relativity with a gravitational Chern-Simons term \cite{Deser:1981wh,Deser:1982vy}.  Including a positive cosmological constant, the action in Lorentzian signature is 
\be
S= {1\over 16 \pi G} \left[\int_{\M} d^3 x \sqrt{-g} \left(R - {2\over \ell^2}\right) + {\mu^{-1}} I_{CS}\right]~,
\ee
where $I_{CS}$ is the gravitational Chern-Simons term
\be
I_{CS} ={1\over 2} \int d^3 x\sqrt{-g} \epsilon^{\lambda \mu\nu}\Gamma^\tau_{\lambda\sigma}
\left(
\partial_\mu \Gamma^\sigma_{\tau \nu} +{2\over 3} \Gamma^\sigma_{\mu\alpha}\Gamma^\alpha_{\nu\tau}
\right)~.
\ee
Here $\mu$ is a real coupling constant with dimensions of mass.  The equations of motion of this theory are third order in derivatives of the metric, so unlike  three dimensional Einstein gravity this theory possesses a propagating local degree of freedom.

In Euclidean signature the action is
\be\label{ICS}
S= {1\over 16 \pi G} \left[\int_{\M} d^3 x \sqrt{g} \left(R - {2\over \ell^2}\right) + {i \mu^{-1}} I_{CS}\right]~,
\ee
where $I_{CS}$ is now the Euclidean Chern-Simons term
\be
I_{CS} ={1\over 2} \int d^3 x\sqrt{g} \epsilon^{\lambda \mu\nu}\Gamma^\tau_{\lambda\sigma}
\left(
\partial_\mu \Gamma^\sigma_{\tau \nu} +{2\over 3} \Gamma^\sigma_{\mu\alpha}\Gamma^\alpha_{\nu\tau}
\right)~.
\ee
We emphasize the appearance of the factor of $i$ in \rref{ICS}.  This arises because the Chern-Simons Lagrangian  is odd under time reversal $t \to -t$, so picks up a factor if $i$ under the Wick rotation $t \to it $.  Thus, as is usually the case for parity non-invariant theories, the action is complex in Euclidean signature.  We will use units where $\ell=1$ and define the dimensionless coupling $k=\ell/4 G$.

The equations of motion are found by varying \rref{ICS} with respect to the metric.  The Euclidean signature equations of motion are
\be\label{CSeom}
G_{\mu\nu} + g_{\mu\nu} = {i \mu^{-1}} C_{\mu\nu}~,
\ee
where $C_{\mu\nu}$ is the Cotton tensor $C_{\mu\nu} = \epsilon^{\alpha \beta}_{\phantom{\alpha\beta}(\mu} G^{}_{\nu)\beta;\alpha}$.  In this paper we will restrict our attention to real solutions to the equations of motion.\footnote{This represents a choice in our definition of the path integral as a sum over real metrics.  This choice is justified by the fact that, as we will see later, the resulting partition function is convergent.  However, other choices of integration contour through the space of metrics may be possible.  This is the case in Chern-Simons gauge theory \cite{Witten:2010cx} so it would be reasonable to investigate a similar possibility here.  
}  For a real metric, the  left and right hand sides of (\ref{CSeom}) are purely real and purely imaginary, respectively.  Thus they must vanish independently.  In particular, the metric must obey the equation of motion of Einstein gravity with a positive cosmological constant \be \label{eeq}
G_{\mu\nu} + g_{\mu\nu} = 0~.
\ee
When this equation is satisfied  the right hand side of equation \rref{CSeom} will vanish automatically.
We conclude that in Euclidean signature the equations of motion reduce to those of general relativity without a gravitational Chern-Simons term.\footnote{This argument was made in the case of TMG with a negative cosmological constant in \cite{Maloney:2009ck}.} 

It is now straightforward to enumerate the smooth solutions to the equations of motion.  Equation \rref{eeq} states that the metric must be locally $S^3$. 
The classification of locally spherical geometries is a standard part of the classification of three-manifolds; see e.g. \cite{thurston97} for a review. We will simply summarize the results here.
The solutions to the equations of motion are the three-manifolds $S^3/\Gamma$, where $\Gamma\subset SO(4)$ is a freely acting discrete subgroup of the isometry group of the sphere.  There are an infinite number of possible subgroups $\Gamma$, which can be enumerated explicitly; they are central extensions of the crystallographic point groups in three dimensions.

As discussed in \cite{Castro:2011xb}, there is a special class of solutions to the Euclidean equations of motion which have a natural physical interpretation.  These are the lens spaces $L(p,q)$, which are quotients of the three sphere by the cyclic group $S^3/\Z_p$.  These spaces are the positive cosmological constant analogue of the BTZ black hole solutions with negative cosmological constant \cite{Park:1998qk}.  They have a straightforward Lorentzian interpretation which we now  review. 

We start by considering the physics of a timelike observer in de Sitter space.  This observer is in causal contact with the static patch of de Sitter space, which has metric
\be
ds^2= dr^2-\cos^2r dt^2+\sin^2r d\phi^2~, \quad \phi\sim \phi + 2\pi n~, \quad \forall n \in \Z~.
\ee
The Euclidean geometry is obtained by taking
\be
t\to t_E=it~,
\ee
which gives the  metric
\be\label{metriclens}
ds^2 = dr^2 + \cos^2 r dt_E^2 + \sin^2 r d\phi^2 ~.
\ee
The geometry has to be smooth at $r=\pi/2$, implying that the Euclidean  time coordinate $t_E$ must be periodically identified. Thus
\be\label{sph}
(t_E,\phi) \sim (t_E,\phi) + 2\pi (m,n) ~~~~~ \forall n,m \in \Z~,
\ee
The geometry \rref{metriclens} with the identifications \rref{sph} is the sphere $S^3$. 

However, there are other identifications of the $t$ and $\phi$ coordinates which make the geometry \eqref{metriclens} smooth. In particular
\be\label{ldef}
(t_E,\phi) \sim (t_E,\phi) + 2\pi \left({m\over p},m{q\over p} + n\right) ~~~~~ \forall n,m \in \Z~,
\ee
is a smooth geometry provided $p$ and $q$ are relatively prime integers.  These identifications define  the lens space $L(p,q)=S^3/\Z_p$; the sphere $S^3$ is $L(1,0)$. We note that a shift of $q$ by a multiple of $p$ can be absorbed into a shift of $n$ in \rref{ldef}.  Therefore the parameter $q$ can be taken to be between $1$ and $p$.   

To understand the physics of these geometries, we note that the operator that generates the identification \eqref{ldef} is
\be\label{rhois}
\rho=\exp\left(-{2\pi\over p}H-{2\pi\over p}iq J \right)~,
\ee
where the charges $H$ and $J$ generate time translation and rotation, respectively. 
Equation \rref{rhois} can be regarded as a density matrix which defines a grand canonical ensemble with temperature $\beta=2\pi/p$ and angular potential $\theta={2\pi}iq/p$.  

This provides a natural physical interpretation for the lens space $L(p,q)$.  At the level of quantum field theory in a fixed de Sitter background, correlation functions in $L(p,q)$ can be Wick rotated to obtain correlation functions in de Sitter which are evaluated in a thermal state of fixed temperature and angular potential.  Once gravitational effects are included, each $L(p,q)$ represents a contribution to the Hartle-Hawking state.  The dominant contribution is from the sphere $S^3=L(1,0)$, which has vanishing angular potential.  This contribution describes the standard thermal behaviour of de Sitter space.  The other lens spaces give subleading contributions which lead to deviations from thermality.  The partition function is then the norm of this Hartle-Hawking state.   

We emphasize that the lens spaces are the only Euclidean geometries with  a clear Lorentzian interpretation. The other saddles of the form $S^3/\Gamma$, where $\Gamma$ is not a cyclic group, can not be Wick rotated to the static patch of de Sitter space.  We will therefore focus in what follows on the sum over lens spaces.  The sum over other saddles can be included, but we do not expect that this will lead to qualitatively different results.  This was discussed in detail in \cite{Castro:2011xb}. 

\subsection{The Gravitational Chern-Simons Action}\label{sec:TMG}

We now need to compute the classical action of one of the $L(p,q)$ saddles, including the effect of the gravitational Chern-Simons term.  This is most easily done using the Chern-Simons formulation of three dimensional gravity, where the action is expressed not as a function of the metric but rather as a function of the frame fields $e^a$ and the connection $\omega^{a}$.  We will follow the same conventions as  \cite{Castro:2011xb}. 

We first note that the Chern-Simons formulation of TMG is somewhat more subtle than that of Einstein gravity.  In the Einstein gravity case, the equations of motion are completely equivalent to those of a Chern-Simons theory; this is implied by the famous equivalence between the second order (metric) and first order (Palatini) formulations of general relativity.  In the case of TMG, however, the equations of motion are third order in the metric and describe a propagating local degree of freedom.  This local degree of freedom is not present in the Chern-Simons formulation; Chern-Simons theory is topological.  So the two theories are not equivalent, even classically.

Nevertheless, the Chern-Simons formulation can be used to evaluate the classical action of TMG for certain solutions.  In particular, if the  Cotton tensor vanishes  (i.e. we are studying a solution of Einstein gravity) then the TMG action is precisely that of a Chern-Simons gauge theory.   This is exactly what happens when we restrict our attention to real metrics in Euclidean signature.

Explicitly, the Chern-Simons action is
\be\label{csaction}
I[A] =  \int_{\M} {\rm Tr}\left( A \wedge dA + {2\over 3} A\wedge A\wedge A \right)~,
\ee
where $A$ is an $SU(2)$ connection and ${\rm Tr}$ is the usual trace on the $SU(2)$ Lie algebra. When the Cotton tensor vanishes, the TMG action \eqref{ICS} can be written as that of an $SU(2)\times SU(2)$ Chern-Simons theory
\be\label{css}
S = - \left( {ik_+\over 4\pi} I[A_+] + {ik_{-}\over 4\pi} I[A_-]\right)~,
\ee
The gauge fields $A_\pm$ are related to the frame fields $e^a$ and spin connection $\omega^a={1\over 2}\epsilon^a{}_b{}^c \omega^b{}_c$ by
\be
A^a_\pm = \omega^a \pm e^a ~,\quad  A_\pm = A^a_\pm T_a~.
\ee
Here $T^a$ are the $SU(2)$ generators.  The levels $k_\pm$ are complex and are related to the gravitational couplings by
\be
i(k_+ - k_-) = {\ell \over 2 G}\equiv  2k ,~~~~k_++k_-= {2k \over \ell \mu}~.
\ee
We now use this to compute the action of a lens space.

Using the metric \eqref{metriclens}, the $SU(2)\times SU(2)$ connection on $L(p,q)$  is
\be\label{Ais}
A_\pm = A_\pm^a T_a=\pm T_1\, dr + ( \cos r \,T_2\pm \sin r\, T_3) d\theta_{\pm}~,
\ee
where $\theta_\pm = \phi\pm t_E$ obey the identifications
\be\label{thetais}
\theta_\pm \sim \theta_\pm + 2 \pi {n (q\pm1) - m p\over p}~, ~~~~~\forall n,m\in \Z~.
\ee
We note that the lens space $L(p,q)=S^3/\Z_p$ contains a non-contractible cycle given by the $\Z_p$ quotient; this is the cycle $(n,m)=(1,0)$ in \rref{thetais}. 
To describe this connection more geometrically, we should compute its holonomy around this non-contractible cycle.  As $\pi_1(L(p,q))=\Z_p$, this holonomy is a $p^{th}$ root of unity in $SU(2) \times SU(2)$ so is conjugate to an element 
\be
\left(\left({e^{2 \pi i n_+/p}~~0 \atop 0~~ e^{-2\pi i n_+/p}}\right),\left({e^{2 \pi i n_-/p}~~0 \atop 0 ~~e^{-2\pi i n_-/p}}\right)\right)\in SU(2)\times SU(2)
\ee  
It is straightforward to compute the holonomy of the connection \rref{Ais} and show that it is given by 
\be\label{hol}
(n_+,n_-) = \left({q + 1 \over 2},{q-1\over 2}\right)~.
\ee

We now compute the action of TMG using known expressions for the $SU(2)$ Chern-Simons invariant on a Lens space.
 For an $SU(2)$ connection on a lens space with holonomy $n$, the Chern-Simons invariant is \cite{Jeffrey:1992tk}
\be\label{inv}
 {1\over 8\pi^2}\int {\rm Tr}\left( A \wedge dA + {2\over 3} A\wedge A\wedge A  \right)= {q^* \over p}n^2~,
\ee
where $q^*$ is the inverse of $q$ mod $p$:
\be
q^*q=1\,{\rm mod}\,p~.
\ee
From \eqref{hol} and \eqref{inv} we find that the action of TMG is
\bea\label{tmg}
Z_{(p,q)}^{(0)} &=& \exp\left({ik_{+}\over 4\pi} I[A_+] + {ik_{-}\over 4\pi}I[A_-]\right)\cr&=&
\exp\left(\pi i(k_{+} - k_{-}) {q^*q\over p} + \pi i (k_{+} + k_{-}){(q^2+1)q^*\over 2p}\right)\cr
&=& \exp\left({2\pi k \over p} + \pi i {k\over \mu} {(q+q^*)\over p}\right)~.
\eea
The first term in this action is real; it is simply the usual Einstein action.  This term is proportional to the volume of the lens space $L(p,q)=S^3/\Z_p$.  The second term is purely imaginary and comes from the gravitational Chern-Simons term.

We note that in general the action of a Chern-Simons gauge theory is invariant under large gauge transformations only if the real part of the Chern-Simons coupling (i.e. the level) is an integer.  In the present case this implies that the gravitational Chern-Simons coupling must be quantized\footnote{We note that a factor of two appears in this expression because the  Chern-Simons gauge group corresponding to three dimensional gravity not actually $SU(2)\times SU(2)$ but rather $SO(4) = SU(2)\times SU(2) / \Z_2$.} 
\be\label{muq}
{k \over 2 \mu} \in \Z~.
\ee
In the gravitational language this condition is necessary for the action of TMG to be invariant under large diffeomorphisms in Euclidean signature.  From now on we will demand that $\mu$ is quantized in accordance with \rref{muq}. Indeed, it is only with this quantization that the action \rref{tmg} is invariant under shifts of $q$ or $q^*$ by a multiple of $p$.

\subsection{The Sum over Geometries}

We  now compute the sum over geometries including the tree level contributions described above.

The sum over lens spaces is
\bea
Z_{\rm lens}&=&\sum_{p=1}^\infty\sum_{(p,q)=1} Z_{(p,q)}^{(0)}\cr
&=& \sum_{p=1}^\infty e^{2\pi k/p}S(k/(2\mu),k/(2\mu),p)~.
\eea
Here $S(a,b,m)$ is the Kloosterman sum
\be
S(a,b,m)=\sum_{n=1\atop (n,m)=1}^m\exp(2\pi i(an^*+bn)/m)~,
\ee
where $n^*$ is the inverse of $n$ mod $m$.
Expanding the exponential we find
\be\label{ZZ}
Z_{\rm lens}= \sum_{r=0}^\infty{(2 \pi k)^r\over r!} \left(\sum_{p=1}^\infty p^{-r} S(k/(2\mu),k/(2\mu),p)\right)~.
\ee
The quantity in parenthesis defines what is known as the Kloosterman zeta function, which appears frequently in number theory.  We include in appendix \ref{ap:sum} a summary of various features of this zeta function. 
This zeta function has a pole at $r=0$, implying that  the sum over tree level geometries diverges.

We note that the divergence described above is not of particular physical significance, as one-loop effects can (and will) change the $p$ dependence of the terms in the sum \rref{ZZ}.  However, it is useful at this point to note that the the sum \rref{ZZ} is already considerably more convergent than the corresponding sum in Einstein gravity.  In that case all of the $r\le 2$ terms diverged  \cite{Castro:2011xb}.  The improved convergence comes from the phases which appeared in the action \rref{tmg}, which lead to cancelations in the sum over geometries.  These cancelations appear because the phase of the Kloosterman sum is essentially randomly distributed.\footnote{In fact, the distribution of the values of the Kloosterman sum is a topic of great interest to number theorists \cite{Sarnak:2000, Sarnak:2009}.} 

We now turn to the discussion of quantum corrections, which will  render this sum finite.  

\section{Quantum Corrections}

In this section we compute the one-loop quantum corrections to the TMG partition function by evaluating the appropriate functional determinants.   We will then perform the sum over geometries including quantum corrections and demonstrate that the result is finite.  Technical details related to the computation of  one-loop determinants are included in the appendices.

\subsection{One loop determinant}

The one loop contribution $S^1$ to the partition function
\be
Z=\int {{\cal D }g\over V_{\rm diff}}e^{-S[g]}=\sum_{g_c}e^{-S^0+S^1+\ldots}~,
\ee
is obtained by integrating over the linearized fluctuations $h$ around each classical saddle. 
The computation is rather similar to that in Einstein gravity, the only difference being that we now have an additional propagating degree of freedom.  

A detailed derivation of the functional determinants appearing in $S^1$ is provided in appendix \ref{aa:tmg}.  This involves a careful accounting of the residual gauge symmetry and Fadeev-Popov ghosts in the measure factor. 
Related discussions in the case of Einstein gravity appear in \cite{Gibbons:1978ji,Christensen:1979iy,Yasuda:1983hk} and for topologically massive gravity with a negative cosmological constant in \cite{Gaberdiel:2010xv}. 

The answer is the following product of functional determinants 
\bea\label{one}
Z^{(1)}=e^{S^{1}}=D_{zm}\frac{\det'^{1/2}(-\Delta_{(1)}-2)_T}{\det'^{1/2}(1+i\mu^{-1} D_M)_{TT}\det'^{1/2}(-\Delta_{(2)}+2)_{TT}}~.\eea
Here  $\Delta_{(j)}=\nabla^\alpha\nabla_\alpha$ is the Laplacian acting on a field of spin $j$ and the subscript $T$ ($TT$) refers to the functional determinant on the space of  transverse vectors (transverse traceless tensors).  The prime indicates that we restrict to the positive part of the spectrum and $D_{zm}$ denotes the contribution from zero  modes.
The operator $D_M$ acts on symmetric 2-tensors and is defined by
\be\label{dm1}
D_M T_{\mu\nu}={1\over 2}(\epsilon_{\mu}^{~\alpha\beta}\nabla_{\alpha}T_{\nu\beta}+\epsilon_{\nu}^{~\alpha\beta}\nabla_{\alpha}T_{\mu\beta})~.
\ee

As one might expect, this formula is closely related to that of Einstein gravity.  The only difference is the determinant involving $D_M$, which comes from the local degree of freedom.  Thus we can write \rref{one} as
\bea
Z^{(1)}&\equiv&{Z_{\rm Ein}^{(1)} Z_{MG}^{(1)}}~,
\eea
where $Z_{\rm Ein}^{(1)}$ is the one-loop determinant for Einstein gravity and
\bea\label{ax}
Z_{MG}^{(1)}= \det {}^{-1/2}(1+i\mu^{-1} D_M)_{TT}~.
\eea
We may now use the results of \cite{Castro:2011xb}, where $Z_{\rm Ein}^{(1)}$ was evaluated explicitly.
For a lens space $L(p,q)$ with $q\neq \pm 1 \mod p$ we have
\be\label{Z1lens}
Z^{(1)}_{\rm Ein, lens}={2\pi\over kp}\left[\cos\left({2\pi\over  p}\right)-
\cos\left({2\pi q\over  p}\right)\right]
\left[\cos\left({2\pi\over  p}\right)-
\cos\left({2\pi q^*\over  p}\right)\right]~,
\ee
 and 
 \be\label{Zq1}
Z^{(1)}_{{\rm Ein},(p,1)}=Z^{(1)}_{{\rm Ein},(p,p-1)}={\pi\over 2kp^2}\sin^2\left({2\pi\over  p}\right)~,
\ee
for $p\neq2$.  Finally, for $S^3$ and $L(2,1)$ we have
 \be\label{Z21a}
 Z^{(1)}_{{\rm Ein}, S^3}={\pi^3\over 2^5k}~,\quad Z^{(1)}_{{\rm Ein}, (2,1)}={\pi^3\over 2^{11}k}~.
 \ee

We now turn to the computation of $Z_{MG}^{(1)}$. 

\subsection{Massive graviton determinant}

We need to compute the one loop determinant for the massive mode 
\bea\label{axx}
Z_{MG}^{(1)}= \det {}^{-1/2}(1+i\mu^{-1} D_M)_{TT}~.
\eea
We will use heat kernel techniques to compute the spectrum of this operator, and zeta function regularization to compute the functional determinant. This regularization proceeds as follows.  The operator $D_M$ has discrete eigenvalues $\lambda_n$ with degeneracies $d_n$ which we will compute explicitly below.  The logarithm of the functional determinant is
\be\label{logz}
\log Z_{MG}^{(1)}= -{1\over2}\sum_n d_n \ln(1+i\mu^{-1}\lambda_n)~.
\ee
This can be written in terms of the zeta function
\be\label{zMG}
\zeta_{MG}(s)=\sum_n {d_n\over (1+i\mu^{-1}\lambda_n)^s}~,
\ee
as
\be\label{logZ}
\log Z_{MG}^{(1)}={1\over 2}{d\over ds}\zeta_{MG}(0)~.
\ee
The functional determinant is formally divergent, but by analytically continuing
 $\zeta_{MG}(s)$ to the whole complex $s$ plane and evaluating \rref{logZ} at $s=0$ we obtain a regulated answer for the determinant.

We now need to construct the zeta function \eqref{zMG} by identifying the eigenvalues and  degeneracies of $D_M$. The first observation is that the square of $D_M$ is the standard  laplacian when acting on symmetric, transverse traceless tensors
\be
D_M^2 T_{\mu\nu}^{\:(TT)}=(-\Delta_{(2)} +3)T_{\mu\nu}^{\:(TT)}~.
\ee
This follows directly from \eqref{dm1}.  This identity relates the spectrum of $\Delta_{(2)}$ with the spectrum of $D_M$. If the operator $-\Delta_{(2)}$ has eigenvalues $\alpha_n$ with degeneracies $d_n$, then $D_M$ has eigenvalues $\lambda_n=\sqrt{\alpha_n+3}$ with the same degeneracy $d_n$.\footnote{In principle there is an ambiguity in the sign of $\lambda_n$, since we could take either branch of the square root.  Because $D_M$ is a self-adjoint operator its spectrum is bounded, so the sign of $\lambda_n$ is fixed for all but a finite number of eigenvalues.  This amounts to an ambiguity of the phase of the determinant $\det(1+ i \mu^{-1}D_M)^{1/2}$ coming from this finite product of eigenvalues.   At the end of the day we will only be interested in the norm of the determinant, so this ambiguity is irrelevant for our purposes.}

Using this relation between $D_M$ and $\Delta_{(2)}$ we can build $\zeta_{MG}(s)$. From \cite{David:2009xg} we know the spectrum of the spin-2 Laplacian on $L(p,q)$. The eigenvalues and degeneracies of the transverse-traceless modes are
\be
\alpha_n=(n+3)^2-3 ~
\ee
and 
\be\label{deg}
d_n={1\over p}\sum_{m\in \Z_p}\chi_{({n\over 2})}(m\tau)\chi_{({n\over 2}+2)}(m\bar \tau)+\chi_{({n\over 2}+2)}(m\tau)\chi_{({n\over 2})}(m\bar \tau)~,
\ee
Here $n=0,1,\ldots$ and we have defined %
\be
\tau=\tau_1-\tau_2= {2\pi \over p}(q-1) ~,\quad \bar\tau=\tau_1+\tau_2= {2\pi \over p}(q+1)~
\ee
and 
\be
\chi_{n}(\tau)={\sin((2n+1){\tau\over 2})\over \sin({\tau\over 2})}~.
\ee
Gathering the above results the zeta function \eqref{zMG} becomes
\be\label{zmg2}
\zeta_{MG}(s)=\sum_{n=0}^\infty d_n(1+i\mu^{-1}(n+3))^{-s}~.
\ee

We now wish to understand the analytic properties of \eqref{zmg2}. Even though it is difficult to evaluate explicitly equation \rref{deg} for the coefficients $d_n$ we note that the sum can be re-arranged in terms of Hurwitz zeta functions.  This follows from the fact that  the coefficients $d_n$  are almost periodic
\be
 d_{rp +j}=d_{j}+2r(rp+2j+6)~,
\ee
 for $q\neq \pm 1 \mod p$.  Special values need to be discussed separately, which we will do in appendix \ref{aa:num}.  This allows us to write \eqref{zmg2} as
\bea\label{zfinal}
\zeta_{MG}(s)&=&\left({\mu \over ip}\right)^s\sum_{j\in \Z_p} \left[d_j-{2p}|a_j|^2\right]\zeta\left(s,a_j\right)\cr && +{2p}\left({\mu\over ip}\right)^{s}\sum_{j\in\Z_p}\left[\zeta(s-2,a_j)+{2\over p}i\mu\zeta(s-1,a_j)\right]~,
\eea
where
\be\label{defa}
a_j={1\over p}(j+3-i\mu)~.
\ee
The advantage of this expression is that we know the analytic properties of the Hurwitz function $\zeta(s,a)$ and its derivatives with respect to $s$. Using \eqref{hurwitz} and \eqref{zfinal} we write \eqref{logZ} as
\bea\label{xx}
\log Z_{MG}^{(1)}&=& N\ln\left({\mu\over i p}\right) +{1\over 2}\sum_{j\in \Z_p}\left(d_j-{2p}|a_j|^2\right)[\ln\Gamma\left(a_j\right)-{1\over 2}\ln (2\pi)] \cr &&+p\sum_{j\in\Z_p}\left[\zeta'(-2,a_j)+{2\over p}i\mu\zeta'(-1,a_j)\right]~.
\eea
It is also useful to compute the norm of $Z_{MG}^{(1)}$, which is\footnote{This is a variant of  the Chowla-Selberg formula. If $(d_j-2p|a_j|^2)$ is of order one we can easily compute the product of gamma functions using Gauss's multiplication formula.}
\bea\label{Z_MG}
|Z_{MG}^{(1)}|^2=(2\pi)^{-A/2}e^{\pi {\rm Im}(N)+B}\left(\mu\over p\right)^{{2\rm Re }(N)}\, \prod_{j=0}^{p-1}\left|\Gamma\left({j+3-i\mu\over p}\right)\right|^{(d_j-{2p}|a_j|^2)}~.
\eea
In \eqref{xx} and \eqref{Z_MG}  we have introduced
\bea\label{functdef}
N&=&{1\over 4} \sum_{j\in\Z_p}d_j(1-2a_j)+{p\over 6}\sum_{j\in\Z_p}\left(a_j^2(a_j+3\bar{a}_j)-{j+3\over p}\right)~,\cr
A&=&\sum_{j\in \Z_p}\left(d_j-{2p}|a_j|^2\right)~,\cr
B&=&2p\sum_{j\in\Z_p}\left[{\rm Re}\zeta'(-2,a_j)-{2\over p}\mu\,{\rm Im}\zeta'(-1,a_j)\right]~.
\eea

\subsection{The Sum over Geometries}

We are now in a position to compute the sum over geometries, including the quantum correction computed above.
The partition function is
\be\label{pitot}
Z= \sum_{(p,q)=1}Z_{(p,q)}^{(0)}Z^{(1)}_{(p,q)} ~,
\ee
where $Z_{(p,q)}^{(0)}$ is given by \eqref{tmg} and the one-loop corrections are
\be
Z^{(1)}_{(p,q)} =Z_{\rm Ein}^{(1)}Z_{MG}^{(1)} ~,
\ee
which are given in \eqref{Z1lens}, \eqref{Zq1}, \eqref{Z21a} and \eqref{xx}. 

Our claim is that \eqref{pitot} is finite, unlike the Einstein gravity case.  To see this, we first investigate the sum excluding the one-loop contribution of the massive graviton $Z_{MG}^{(1)}$.  The sum includes various terms (such as those coming from Lens spaces with $q=\pm 1$ mod $p$) which are finite when summed over $p$.  The only term which is not finite comes from \eqref{Z1lens} and is
\bea\label{ztest}
 \sum_{(p,q)=1}Z_{(p,q)}^{(0)}Z^{(1)}_{\rm Ein} &=& \sum_{p=1}^\infty {2\pi\over k p}e^{2\pi k/p}S(k/(2\mu),k/(2\mu),p) +\ldots~
 \cr &=& k^{-1} \sum_{r=0}^\infty {\left(2 \pi k\right)^{r}\over r!} \left(\sum_{p=1}^\infty { p^{-(1+r)}} S(k/(2\mu),k/(2\mu),p)\right)~+ \ldots
\eea
Again we have encountered the Kloosterman zeta function, which is the quantity in parenthesis in the second line.  As described in appendix C, this zeta function is finite for every value of $r>0$.  The only potentially problematic term is the one with $r=0$, where we are considering the sum
\be\label{zetas}
\sum_{p=1}^\infty p^{-2s} S(k/(2\mu),k/(2\mu),p)~
\ee  
with $s=1/2$.  This particular value of the Kloosterman zeta function is of considerable mathematical interest, as the behaviour of the Kloosterman zeta function on the line $\Re s=1/2$ is related to the spectrum of the hyperbolic laplacian on the modular surface ${\cal H}/SL(2,\Z)$.  In general,  the behaviour of sum \rref{zetas} at $s=1/2$ is a difficult number theoretic question; we refer the reader to \cite{Iwaniec:2002aa, Sarnak:2009} for more details.

In the above discussion, we have neglected the important $Z_{MG}^{(1)}$ term.  To include the effects of this term we must understand its $p$ dependence, which is not immediately evident from  \eqref{xx}.   In order to demonstrate that the partition function is finite it is  sufficient to argue that for large values of $p$ the determinant $Z_{MG}^{(1)}$ decreases sufficiently quickly as a function of $p$.  This can be checked numerically, as we describe in appendix \ref{aa:num}.\footnote{Although we give only numeric evidence in appendix B, we suspect that an analytic proof is possible.} In particular we note that for every value of $\mu$ there is an $\epsilon>0$  such that
\be\label{bound}
|Z_{MG}^{(1)}|^2\leq e^{\pi {\rm Im}(N)+B}\left(\mu\over p\right)^{{2\rm Re }(N)}\, \prod_{j=0}^{p-1}\left|\Gamma\left({j+3\over p}\right)\right|^{(d_j-{2p}|a_j|^2)}={\cal O} (p^{-\epsilon}) ~,
\ee
for sufficiently large $p$.  Even without analytic expression for $Z_{MG}^{(1)}$ as a function of $p$ and $q$, the upper bound \eqref{bound} is sufficient for our purposes.  Armed with this result we can now reconsider the troublesome $r=0$ term in equation \rref{ztest}.  In this case we are now considering a zeta function of the form \rref{zetas} at $s=1/2 + \epsilon$ for some positive value of $\epsilon$.  This sum is convergent.  
We conclude that the partition function of topologically massive gravity in de Sitter space is finite.

\subsection*{Acknowledgments}
We thank D. Anninos, T. Anous, S. Carlip, M. Gaberdiel, S. Giombi, A. Strominger, W. Song and X. Yin for useful conversations.  This work is supported by the National Science and Engineering Council of Canada.

\appendix

\section{Perturbative analysis of TMG}\label{aa:tmg}

In this appendix we construct explicitly the one-loop determinant of TMG. We start by expanding the metric around a background solution $\bar g$ of the equations of motion
\begin{equation}\label{decomp}
 g_{\mu\nu}=\bar{g}_{\mu\nu}+h_{\mu\nu}~,
\end{equation}
where  $h$ is a linearized metric fluctuation. The action takes the form
\be
S[\bar{g} + h]=S^{0}[\bar{g}]+S^{(2)}_{\rm bulk}[h]+S_{\rm gauge}[h]+S_{\rm ghost}+{\cal O}(h^3)~.
\ee
Here $S^{0}$ is the on-shell action, $S^{(2)}_{\rm bulk}$ is the action \eqref{ICS} expanded to quadratic order in $h$ and $S_{\rm gauge}$ and $S_{\rm ghost}$ are the contributions from the gauge fixing terms and ghost determinant, which will be computed explicitly in the next subsection using BRST quantization.  

To obtain $S^{(2)}_{\rm bulk}$ it is convenient to decompose the metric perturbation into its scalar trace and traceless tensor parts 
\begin{equation}
 h_{\mu\nu}=\phi_{\mu\nu}+\frac{1}{3}\bar{g}_{\mu\nu}\phi~,
\end{equation}
with $\phi^\alpha_{\:\:\alpha}=0$.\footnote{In this section indices are raised and lowered with $\bar{g}_{\mu\nu}$ and all covariant derivatives are with respect to the background metric.}  The curvature tensor of the background metric satisfies
\be\label{RR}
\bar{R}_{\mu\alpha\nu\beta}={\bar{R}\over 6}(\bar{g}_{\mu\nu}\bar{g}_{\alpha\beta}-\bar{g}_{\mu\beta}\bar{g}_{\nu\alpha}) ~,\quad \bar{R}_{\mu\nu}=2\bar{g}_{\mu\nu} ~, \quad \bar{R}=6~,
\ee
so the quadratic action can be written as
\bea\label{s2}
 &&S^{(2)}_{\rm bulk}=-\frac{1}{32\pi G}\int d^3x\sqrt{\bar{g}}\left[-\frac{1}{2}\phi^{\mu\nu}(1+i\mu^{-1}D_M)(\Delta_{(2)}+4)\phi_{\mu\nu}+\frac{1}{9}\phi(\Delta_{(0)}+3)\phi\right.\nn\\
&&\left.+\frac{1}{2}\phi^{\mu\nu}(1+i\mu^{-1}D_M)\chi_{\mu\nu}-\frac{1}{3}\phi^{\mu\nu}\theta_{\mu\nu}\right]~.
% &&-\frac{i}{32\pi G\mu}\int d^3x\sqrt{\bar{g}}\left[-\frac{\phi^{\mu\nu}}{2}D_M\psi_{\mu\nu}-2\phi^{\mu\nu}D_M\phi_{\mu\nu}+\frac{\phi^{\mu\nu}}{2}D_M\chi_{\mu\nu}\right]
\eea
Here $\Delta_{(j)}=\nabla^\alpha\nabla_\alpha$ is the Laplacian acting on a field of spin $j$ and 
we have introduced the operator $D_M$ 
\be\label{dm}
D_M T_{\mu\nu}={1\over 2}(\epsilon_{\mu}^{~\alpha\beta}\nabla_{\alpha}T_{\nu\beta}+\epsilon_{\nu}^{~\alpha\beta}\nabla_{\alpha}T_{\mu\beta})~,
\ee
which acts on symmetric 2-tensors.  We have defined the fields
\bea
%  \psi_{\mu\nu}&=&\Delta_{(2)}\phi_{\mu\nu}~,\cr
\chi_{\mu\nu}&=&\nabla^\lambda\nabla_\mu\phi_{\lambda\nu}+\nabla^\lambda\nabla_\nu\phi_{\lambda\mu}~,\cr
\theta_{\mu\nu}&=&\nabla_\mu\nabla_\nu\phi~ .
\end{eqnarray}

\subsection{BRST quantization}

We now turn to the gauge fixing conditions and Fadeev-Popov determinants. For a higher derivative theory such as TMG it is convenient to use the BRST approach, which gives a derivation of the ghost determinants while preserving gauge invariance. For a more detailed discussion of the BRST quantization of Chern-Simons theories and TMG see respectively \cite{BarNatan:1991rn} and \cite{Deser:1990bj}.  

 The BRST transformations are
\begin{eqnarray}\label{BRST}
\delta g_{\mu\nu}&=&\mathcal{L}_cg_{\mu\nu}=c^\alpha \p_\alpha g_{\mu\nu}+g_{\nu\alpha}\p_\mu c^\alpha+g_{\mu\alpha}\p_\nu c^\alpha~,\cr
\delta c^\alpha&=&(\p_\beta c^\alpha)c^\beta~,\cr
\delta \bar{c}^\alpha&=&-B^\alpha~,\cr
\delta B^\alpha&=&0~,
\eea
where $c$ and $\bar{c}$ are the Fadeev-Popov ghost and anti-ghost vector fields and $B$ is an auxiliary field. The notation is chosen to resemble the usual Fadeev-Popov procedure, but we note that $c$ and $\bar{c}$ are not complex conjugates. It is straightforward to check
that $\delta^2=0$. 
The gauge fixed action is
\begin{equation}
 S_{\rm tot}=S+\delta \Psi~,
\end{equation}
where $\Psi$ is an arbitrary functional of ghost number -1. The variation $\delta \Psi$ includes both  $S_{\rm gauge}$ and $S_{\rm ghost}$. The traditional choice for $\Psi$  is
\begin{equation}
\Psi=\frac{-1}{16\pi G}\int d^3x\sqrt{g}\:\bar{c}^\alpha\left(\frac{1}{2}B_\alpha-G_\alpha\right)~,
\end{equation}
where $G_\alpha$ is the gauge fixing condition. One can then solve for $B$ and plug back into the action. However, this procedure gives us a gauge-fixing term
proportional to $G_\alpha G^\alpha$ whereas we need a term third order in derivatives. So we will take the functional $\Psi$ to be
\begin{equation}
 \Psi=\frac{-1}{16\pi G}\int d^3x\sqrt{g}\: \bar{c}^\alpha(D_B)_{\alpha}^{\:\:\beta}\left(\frac{1}{2}B_\beta-G_\beta\right)~,
\end{equation}
where
\begin{equation}
 (D_B)_{\alpha}^{\:\:\beta}=\delta_{\alpha}^{\:\:\beta}+\frac{i}{2\mu}\epsilon_{\alpha}^{\:\:\mu\beta}\nabla_\mu~.
\end{equation}
Since we are interested in the one-loop part of the path-integral, we expand the metric around a saddle point
\begin{equation}
 g_{\mu\nu}=\bar{g}_{\mu\nu}+h_{\mu\nu}~,
\end{equation}
and keep the terms that are at most quadratic in $B$, $c$, $\bar{c}$ and $h_{\mu\nu}$.
We will take the gauge-fixing condition to be
\begin{equation}\label{gauge}
 G_\alpha={\nabla}^{\beta} (h_{\alpha\beta}-\frac{1}{2}\bar{g}_{\alpha\beta}h)~,
\end{equation}
where $h=\bar{g}^{\alpha\beta}h_{\alpha \beta}$. At linear order
\be
 \delta G^{(1)}_\alpha=(g_{\alpha\beta}{\nabla}^2+\bar{R}_{\alpha\beta})c^\beta~,
\ee%
so
\begin{equation}
 \delta\Psi^{(2)}=\frac{-1}{16\pi G}\int d^3x\sqrt{\bar{g}}\left(-B^\alpha(D_B)_{\alpha}^{\:\:\beta}\left(\frac{1}{2}B_\beta-G_\alpha\right)-\bar{c}^\alpha (D_B)_{\alpha}^{\:\:\gamma}(g_{\gamma\beta}\bar{\nabla}^2+\bar{R}_{\gamma\beta})c^\beta\right)~.
\end{equation}
Rewriting the equation above gives
\begin{eqnarray}\label{BRSTfunc}
 \delta\Psi^{(2)}&=&\frac{1}{16\pi G}\int d^3x\sqrt{\bar{g}}\left(-\frac{1}{2}G^\alpha (D_B)_{\alpha\beta}G^\beta+\bar{c}^\alpha (D_B)_{\alpha}^{\:\:\gamma}(g_{\gamma\beta}\bar{\nabla}^2+\bar{R}_{\gamma\beta})c^\beta\right)\nn\\
&&+\frac{1}{16\pi G}\int d^3x\sqrt{\bar{g}}\:\frac{1}{2}(B^\alpha-G^\alpha)(D_B)_{\alpha\beta}(B^\beta-G^\beta)~.
\end{eqnarray}
Using (\ref{gauge}) and (\ref{decomp}) we obtain the following form for the gauge-fixing term
\begin{eqnarray}\label{BRSTact}
 \delta\Psi^{(2)}&=&\frac{1}{32\pi G}\int d^3x\sqrt{\bar g}\Big(-3\phi^{\mu\nu}(1+i\mu^{-1}D_M)\phi_{\mu\nu}+\frac{1}{36}\phi\Delta_{(0)}\phi+\frac{1}{2}\phi^{\mu\nu}(1+i\mu^{-1}D_M)\chi_{\mu\nu}\cr&&-\frac{1}{3}\phi^{\mu\nu}\theta_{\mu\nu} 
% +\frac{i}{2\mu}\phi^{\mu\nu}D_M\chi_{\mu\nu}-\frac{3i}{\mu}\phi^{\mu\nu}D_M\phi_{\mu\nu}
+2\bar{c}^\alpha(D_B)_{\alpha}^{\:\:\gamma}(g_{\gamma\beta}\bar{\nabla}^2+\bar{R}_{\gamma\beta})c^\beta+(B^\alpha-G^\alpha)(D_B)_{\alpha\beta}(B^\beta-G^\beta)~\Big)~.
\end{eqnarray}

\subsection{One-loop Partition Function}

Collecting all the terms quadratic in $h$ from \eqref{RR}, \eqref{s2} and \eqref{BRSTact}, the one-loop effective action is 
\bea\label{S1}
 S^{1}&=&S^{(2)}_{\rm bulk}[h]+S_{\rm gauge}[h]+S_{\rm ghost}\cr
 &=& -\frac{1}{32\pi G}\int d^3x \sqrt{\bar{g}}\left[\frac{1}{2}\phi^{\mu\nu}\left(1+i\mu^{-1}D_M\right)(-\Delta_{(2)}+2)\phi_{\mu\nu}+\frac{1}{12}\phi(\Delta_{(0)}+4)\phi\right]\cr
&&-\frac{1}{16\pi G}\int d^3x \sqrt{\bar{g}}\left[\bar{c}^\alpha (D_B){}_{\alpha}^{\:\:\beta}(-\Delta_{(1)}-2)c^\beta
-\frac{1}{2}(B^\alpha-G^\alpha)(D_B){}_{\alpha\beta}(B^\beta-G^\beta)\right]~.\nn\\
\eea
The one-loop partition function is
\bea\label{zdet}
Z&=&\int \mathcal{D}\phi_{\mu\nu}\mathcal{D}\phi\mathcal{D}c\mathcal{D}\bar{c}\mathcal{D}B\:\:e^{-S^{0}+S^{1}}\nn\\
 &=& e^{-S^{0}}\frac{\det(D_B(-\Delta_{(1)}-2))\det^{-1/2}(D_B)}{\det^{1/2}\left(\left(1+{i}\mu^{-1}D_M\right)(-\Delta_{(2)}+2)\right)\det^{1/2}(-\Delta_{(0)}-4)}~.
\eea
%where $\Delta^{LL}_{(s)}$ is the Lichnerowicz Laplacian acting on an $s$-tensor. Using \eqref{RR}, $\Delta^{LL}_{s}$ is simplified to
%\bea
 %\Delta^{LL}_{(0)}&=&-\Delta_{(0)}\phi~,\cr
 %\Delta^{LL}_{(1)}\phi_\mu&=&-\Delta_{(1)}\phi_\mu+R_{\mu\alpha}\phi^\alpha=(-\Delta_{(1)}+2)\phi_\mu~,\cr
%\Delta_L^{(2)}\phi_{\mu\nu}&=&-\Delta_{(2)}\phi_{\mu\nu}+R_{\mu\alpha}\phi^\alpha_{\:\:\nu}+R_{\nu\alpha}\phi^\alpha_{\:\:\mu}-2R_{\mu\alpha\nu\beta}\phi^{\alpha\beta}=(-\Delta_{(1)}+6)\phi_{\mu\nu}~.
%\eea
One can further simplify this by decomposing these operators into transverse modes:
\begin{eqnarray}
\det((1+i\mu^{-1} D_M)(-\Delta_{(2)}+2))&=&\det((1+i\mu^{-1} D_M)(-\Delta_{(2)}+2))_T\det(D_B(-\Delta_{(1)}-2))\cr
\det(D_B(-\Delta_{(1)}-2))&=&\det(D_B(-\Delta_{(1)}-2))_T\det(-\Delta_{(0)}-4)~.
\end{eqnarray}
 Then,
\begin{equation}\label{detfinal}
Z^{(1)}=e^{S^1}=\frac{\det^{1/2}(D_B(-\Delta_{(1)}-2))_T\det^{-1/2}(D_B)_T}{\det^{1/2}\left(\left(1+i{\mu}^{-1}D_M\right)(-\Delta_{(2)}+2)\right)_{TT}}~,
\end{equation}
where the subscripts denote that the operators act on transverse and traceless tensors.

%$V_{\rm diff}$ into isometries (denoted $V_{\rm KV}$) and gauge fixing conditions (denoted $V_{\rm gauge}$). The measure factor then becomes
%\be\label{pi}
%\int {{\cal D }g \over V_{\rm KV} V_{\rm gauge}}=\int {{\cal D }h\over k\,V_{\rm KV} }\Delta_{\rm ghost}~.
%\ee
%where $\Delta_{ghost}$ is the Fadeev-Popov determinant. 

We now proceed by splitting the determinants appearing in \rref{detfinal} using the formula $\det (AB) = \det(A) \det(B)$.  While this splitting  is straightforward for finite dimensional operators, a subtlety arises for infinite dimensional operators of the sort appearing here.  The problem is that while the expressions $\det (AB)$ and $\det(A)\det(B)$ are equal when regarded as formal products of eigenvalues, the zeta function regularizations of these infinite products may differ.  
The difference between these two expressions is known as a multiplicative anomaly;
see e.g. \cite{Elizalde:1997nd, springerlink:10.1007/BFb0078372} for more details.
%In general, there may be a ``multiplicatively anomaly" defined as
%\begin{equation}
% a(A,B)\equiv\ln\det(AB)-\ln\det(A)\ln\det(B).
%\end{equation}
The operators of interest in (\ref{detfinal}) are all powers of the Laplace operator.  The multiplicative anomaly for Laplace operators has been shown to vanish in odd dimensions \cite{Elizalde:1997nd}.  We will therefore proceed to split the determinants without including a multiplicative anomaly, although this issue may be worthy of further study.

Finally, we note that some of the operators appearing in \rref{detfinal} may have zero or negative eigenvalues.  These arise when one studies geometries which possess  isometries or conformal killing vectors.  To treat the zero modes properly one should not include them in the determinant \rref{detfinal}, but instead integrate directly over the appropriate moduli space of collective coordinates.  The negative modes (which come from the scalar operator $\Delta_{(0)}$) are treated by  rotating the contour of integration in field-space.   Both of these subtleties appear in Einstein gravity and were studied in detail in \cite{Castro:2011xb}; the analysis in TMG is identical, so we will not repeat it here.  

We are left with our final expression for the one-loop determinant
\bea
Z^{(1)}=D_{zm}\frac{\det'^{1/2}(-\Delta_{(1)}-2)_T}{\det'^{1/2}(1+i\mu^{-1} D_M)_{TT}\det'^{1/2}(-\Delta_{(2)}+2)_{TT}}\eea
Here $D_{zm}$ is the zero-mode piece. The prime indicates that we restrict to  positive eigenvalues.

\section{More on the massive graviton}\label{aa:num}

In this appendix we discuss some of the properties of the determinant \eqref{ax}. We will start by evaluating $Z_{MG}^{(1)}$ for special values of $(p,q)$.  In the next subsection we bound the determinant using numerical estimates. 

\subsection{Special cases of $L(p,q)$}

The zeta function associated to $Z_{MG}^{(1)}$ is
\be\label{zmgx}
\zeta_{MG}(s)=\sum_{n=0}^\infty d_n(1+i\mu^{-1}(n+3))^{-s}~.
\ee
with $d_n$ given by \eqref{deg}. 

We start by considering the three-sphere $S^3=L(1,0)$. The degeneracies for this case are
\be\label{degs3}
d_n=2((n+3)^2-4)~.
\ee
This allows us to write \eqref{zmgx} as
\be\label{zs3}
\zeta_{MG, S^3}=2(-i\mu)^{s}\left[\zeta(s-2,-i\mu)+2i\mu\zeta(s-1,-i\mu)-(\mu^2+4)\zeta(s,-i\mu)\right]+{6\over(1+i\mu^{-1})^{s}}
\ee
where we neglect terms independent of $s$. From here it is straightforward to compute $Z_{MG}^{(1)}$ using  \eqref{logZ}. The resulting expression is not illuminating, hence we leave it as an exercise for the curious reader.  

The next geometry that needs separate treatment is $L(2,1)$. Here, the degeneracies are
\be
d_{2n}=2(2n+1)(2n+5)~,\quad d_{2n+1}=0~,
\ee
and the zeta function is
\be
\zeta_{MG, (2,1)}=8\left({\mu\over 2i}\right)^s\left[\zeta\left(s-2,{3-i\mu\over 2}\right)+i\mu\zeta\left(s-1,{3-i\mu\over 2}\right)-(1+{\mu^2\over 4})\zeta\left(s,{3-i\mu\over 2}\right)\right]~.
\ee

\subsection{Numerical estimates of $Z_{MG}^{(1)}$}

Here we numerically explore the $p$-dependence of the expression (\ref{Z_MG}) for the determinant $Z^{(1)}_{MG}$. The claim is that for large values of $p$ -- in particular $p\gg \mu$-- this determinant decreases as a negative power of $p$. Our starting point is the inequality in \eqref{bound}, which reads 
\be\label{aa:bound}
|Z_{MG}^{(1)}|^2\leq e^{\pi {\rm Im}(N)+B}\left(\mu\over p\right)^{{2\rm Re }(N)}\, \prod_{j=0}^{p-1}\left|\Gamma\left({j+3\over p}\right)\right|^{(d_j-{2p}|a_j|^2)}~. 
\ee
The coefficients  $a_j$, $N$ and $B$ are given in \eqref{defa} and \eqref{functdef}.

We first focus on the exponent in the gamma functions.  
\be
\prod_{j=0}^{p-1}\left|\Gamma\left({j+3\over p}\right)\right|^{(d_j-{2p}|a_j|^2)}~. 
\ee
The coefficients $d_j$ for $L(p,q)$ are bounded by the $S^{3}$ values \eqref{degs3}, and from figure \ref{numeTMG} we observe that the maximum value of $d_j$ is of order $p$ which occurs at $j=p-1$.  For sufficiently large values of $p$,  $(d_j-{2p}|a_j|^2)$ is always negative which makes the gamma function in \eqref{aa:bound} negligible.
%negative for most values of $j$ .  

Next, we turn to $N$ which is defined in \eqref{functdef}. The real part of $N$ is
\be
{\rm Re} N= \sum_{j\in\Z_P} \left[d_j+{2\over 3p^2}(j+3)^3-({1\over 6}+{2\over p})(j+3)\right]~,
\ee
which is strictly positive, whereas the imaginary part is
\begin{eqnarray}
 {\rm Im} N=\mu\left(\frac{1}{2p}\sum_j d_j-\frac{p}{3}-\frac{5}{2}-\frac{(2\mu^2+37)}{6p}\right)\lesssim \frac{\mu}{2}~.
\end{eqnarray}
 Therefore the asymptotic behavior of terms in \eqref{aa:bound} that depend on $N$ is 
\be
e^{\pi {\rm Im}(N)}\left(\mu\over p\right)^{{2\rm Re }(N)} \sim e^{\pi\mu/2}\left(\mu\over p\right)^{2|{\rm Re }(N)|}+\ldots~.
\ee

The final term we need to consider is $B$ which is considerably more difficult. Taking $p\gg \mu$ we have
\be\label{aa:B}
B\sim 4p\sum_{j\in\Z_p}\zeta'\left(-2, {j\over p}\right)=\frac{4}{p}\zeta'(-2)\sim -\frac{0.12}{p}~.
\ee
%
% When $j=p$ we have
% %
% \be
% \zeta'(-2,1)=\zeta'(-2)=-{\zeta(3)\over 4\pi^2}<0~,
% \ee
% %
% hence at least one term in \eqref{aa:B} is negative, so we can speculate that $e^B$ is decreasing as a function of $p$. 
Hence $e^B<1$ and can be neglected.

  Gathering our estimates and observations above, we  conclude that for every value of $\mu$ there is a positive number $\epsilon$ such that
  \be\label{ab:bound}
|Z_{MG}^{(1)}|^2\lesssim e^{\pi\mu/2}\left(\mu\over p\right)^{{2\rm Re }(N)}\, \prod_{j=0}^{p-1}\left|\Gamma\left({j+3\over p}\right)\right|^{(d_j-{2p}|a_j|^2)}\sim{\cal O} (p^{-\epsilon}) ~. 
\ee
  We have extensively checked this bound numerically.
  While it is possible to make much stronger estimates for $Z_{MG}$, equation \rref{ab:bound} will be sufficient for our purposes.

%The right hand side can be made yet larger by replacing $d_j$ by its maximum value over all $j$ and $q$:
%\begin{eqnarray}
 %|Z_{MG}^{(1)}|^2&\leq& \left(\frac{\mu}{p}\right)^{{2\rm Re }(N)}\prod_{j=0}^{p-1}\Gamma\left(\frac{j+3}{p}\right)^{\max_{j,q} d_j(p,q)}\nn\\
%&\leq&\left(\frac{\mu}{p}\right)^{{2\rm Re }(N)} \left(\frac{(2\pi)^p}{p^{5/2}}\right)^{\alpha},
%\end{eqnarray}
%where $\alpha=\max_{j,q} d_j(p,q)$ and in the second line we have used Gauss's multiplication formula. 

%Figure 1 shows the exponents $ {\rm Re}(N)$ defined in (\ref{functdef}) and $\alpha$ numerically computed as a function of $p$. The crucial observation is that the maximal degeneracy $\alpha$ grows linearly in $p$ whereas $ {\rm Re}(N)$ appears to grow as $\mathcal{O}(p^{2+\epsilon})$ for positive $\epsilon$. For large $p$ this implies:
%\begin{eqnarray}
 %|Z_{MG}^{(1)}|^2&\sim&\left(\frac{(2\pi)^\beta}{p}\right)^{p^2}\frac{1}{p^{\left(p^\epsilon\right)}}
%\end{eqnarray}
%for $\beta$ and $\epsilon$ positive constants.

\begin{figure}\label{numeTMG}
\begin{center}
\includegraphics[width=0.4\textwidth]{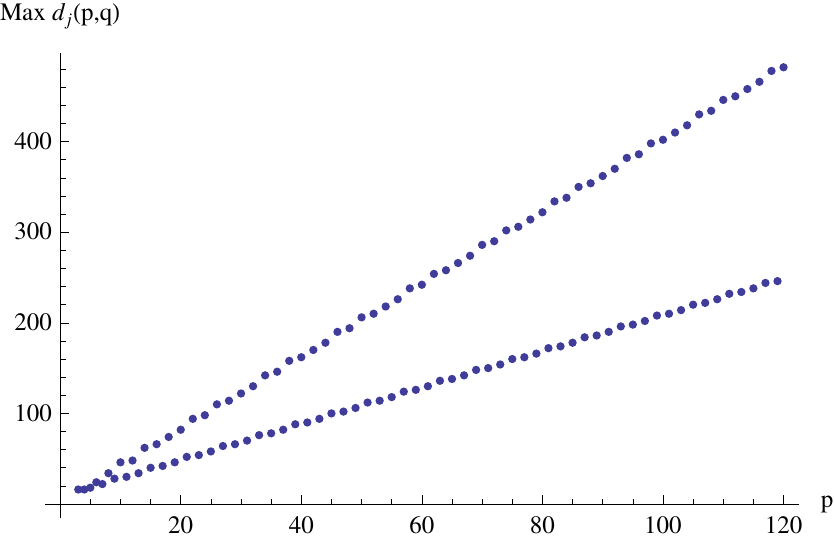}
  \caption{The plot shows the degeneracy of the highest degenerate state as a function of $p$. It suggests a linear growth in $p$ with different slopes for even and odd $p$.}
\end{center}
\end{figure}

\section{Dirichlet Series}\label{ap:sum}

In this appendix we summarize some useful number-theoretic formulae.\\

\noindent{\it Riemann zeta function:} 

The Riemann zeta function $\zeta(s)$ is the analytic continuation of the series 
\be
\zeta(s)=\sum_{n=1}^\infty{1\over n^s}=\prod_{p \,{\rm prime}}(1-p^{s})^{-1} ~,
\ee
to the complex $s$ plane.  
The function has a simple pole at $s=1$ and Laurent series
\be
\zeta(s)={1\over s-1}+\sum_{k=0}^\infty\gamma_k{(-1)^k\over k!}(s-1)^k~,
\ee
where $\gamma_k$ is the Stieltjes constant. Some useful values of $\zeta(s)$ are
\be\label{zeta}
\zeta(0)=-{1\over 2}~,\quad {d\over ds}\zeta(0)=-{1\over 2}\ln (2\pi)~.
\ee

\noindent {\it Hurwitz zeta function:}

A simple generalization of the Riemann zeta function is the Hurwitz function 
\be
\zeta(s,a)=\sum_{n=0}^\infty{1\over (n+a)^s}~.
\ee
It is a meromorphic function in $s$ and ${\Re}(a)>-1$ with a simple pole at $s=1$.  We will need the following values
% the following properties of $\zeta(s,a)$,
%  
\be\label{hurwitz}
\zeta(0,a)={1\over 2}-a~,\quad {d\over ds}\zeta(0,a)=\ln(\Gamma(a))-{1\over 2}\ln (2\pi)~.
\ee
so that in particular
\be\label{gg}
{d\over ds}\zeta(0,a)+{d\over ds}\zeta(0,-a)=-\ln(\sin(\pi a))-\ln(-2a)~.
\ee
%where we used 
%\be\label{gg}
%\Gamma(x)\Gamma(-x)=-{\pi\over x\sin(\pi x)}~.
%\ee
%

\noindent{\it Kloosterman Zeta Function}

We now summarize a few features of  Kloosterman zeta functions. The Kloosterman sum is defined as
\be
S(a,b,m)=\sum_{n=1\atop (n,m)=1}^m\exp(2\pi i(an^*+bn)/m)~,
\ee
where $n^*$ is the inverse of $n$ modulo $m$. 
We are interested in sums of the form 
\be
L(m,n;s) = \sum_{p=1}^\infty p^{-2s} S(m,n;p)~.
\ee
This is known as the Kloosterman zeta function (see \cite{Iwaniec:2002aa} for details).   This series converges absolutely when $\Re s >1/2$.  

The analytic properties are most conveniently summarized by the function 
\be
Z(m,n;s) = {1\over 2 \sqrt{mn}}\sum_{p=1}^\infty p^{-1} S(m,n;p) J_{2s-1} ({4  \pi \over p} \sqrt{mn})~,
\ee
when $mn$ positive, with a similar formula for $mn$ negative.  Using the Neumann expansion
\be
z^{\nu}=2^\nu \sum_{k=0}^\infty {(\nu+2k)\Gamma(\nu+k)\over k!} J_{\nu+2k}(z)~,
\ee
with $z={4  \pi \over p} \sqrt{mn}$ and $\nu=2s-1$
we see that
\be
L(m,n;s) = {2^{2s} \sqrt{mn}\over (4 \pi \sqrt{mn})^{2s-1}}\sum_{k=0}^\infty {(2(s+k)-1)\Gamma(2s-1+k)\over k!}   Z(m,n;s+k)~.
\ee

The only poles of $L(m,n;s)$ on the real $s$ axis come from the gamma function, which has simple poles at the non-positive integers.  For the $k=0$ term in the sum these poles are cancelled by the coefficient $2(s+k)-1$.  Thus $L(m,n;s)$ has no pole at $s=1/2$.  However, when $s=-n/2$, $n=0,1,\dots$ there will be simple poles.  For example, there is a pole at $s=0$ with non-zero residue coming from the $k=1$ term:
\be
L(m,n;s)\sim {1\over s} 4 \pi mn Z(m,n;1) + \dots~.
\ee
Similar conclusions hold for the case where $mn$ is negative.

\bibliographystyle{utphys}
\bibliography{dsref}
\end{document}